# Mid-infrared optical frequency comb generation from a $\chi^{(2)}$ optical superlattice box resonator

Kunpeng Jia[+], Xiaohan Wang[+], Xin Ni, Huaying Liu, Liyun Hao, Jian Guo, Jian Ning, Gang Zhao, Xinjie Lv, Zhenda Xie\*, and Shining Zhu\*

*National Laboratory of Solid State Microstructures, School of Electronic Science and Engineering, College of Engineering and Applied Sciences, and School of Physics, Nanjing University, Nanjing 210093, China*

[+]*These authors contributed equally to this work.*

\* E-mail: xiezhenda@nju.edu.cn,
zhusn@nju.edu.cn

Optical frequency combs (OFCs) at Mid-Infrared (MIR) wavelengths are essential for applications in precise spectroscopy, gas sensing and molecular fingerprinting, because of its revolutionary precision in both wavelength and frequency domain. The microresonator-based OFCs make a further step towards practical applications by including such high precision in a compact and cost-effective package. However, dispersion engineering is still a challenge for the conventional $\chi^{(3)}$ micro-ring resonators and a MIR pump laser is required. Here we develop a different platform of a $\chi^{(2)}$ optical superlattice box resonator to generate MIR OFC by optical parametric down conversion. With near-material-limited quality factor of $2.0 \times 10^7$, broadband MIR OFC can be generated with over 250 *nm* span around 2060 *nm*, where only a common near-infrared laser is necessary as pump. The fine teeth spacing corresponds to a measurable radio frequency beat note at 1.566 *GHz*, and also results in a fine spectroscopy resolution. Its linewidth is measured to be 6.1 *kHz*, which reveals a low comb noise that agrees well with the clean temporal waveforms. With high output power of over 370 *mW*, such MIR OFC is capable for long distance sensing and ranging applications.

Optical frequency comb (OFC) is a ruler in optical frequency, with grid lines formed by inherent pristine frequency spacing, and leads to the revolutionary precision metrology[1-13]. There is special interest to generate OFC at mid-infrared (MIR) wavelengths between 2 and 20 $\mu m$, where molecules spectral fingerprints can be explored for high resolution spectroscopy, sensing and ranging applications. However, broadband MIR OFC is difficult for direct generation, following the conversional approach using mode-locked lasers [14] which is limited by the few choices of MIR laser gain mediums. In contrast, the indirect MIR OFC generation is normally a more effective way. One approach is the frequency extension from an existing near infrared (NIR) OFC, by the means of difference frequency generation[15-19], synchronously pumped optical parametric oscillation[20-24], and supercontinuum generation[25-30]. This approach can effectively transfer the pump coherence to the MIR wavelength but high-quality pump OFCs are required. The second approach is the Kerr comb [8,13,31-33] generation from a continuous wave (CW) MIR pump laser. Taking advantage of the fast development of the microresonators over the past decade [3,5,9,12,34-37], it can be potentially realized in a compact package. However, a stable MIR pump laser is still needed to reach high comb quality. In both approaches, dispersion needs to be perfectly engineered for the MIR wavelengths, which is usually realized by special design of mode profile and precise manufacture. On the other hand, a $\chi^{(2)}$ optical parametric oscillator (OPO) can down convert a NIR laser beam to MIR range, which is potentially good for MIR OFC generation[38-42]. In a domain-engineered optical superlattice [43,44,45], the quasi-phase matching (QPM) of this down conversion process can be independently engineered from the material dispersion. This allows extra flexibility for comb generation while a larger QPM bandwidth is engineered. Such $\chi^{(2)}$ OPO can also be integrated in high quality factor (Q) microresonators, for example, whispering gallery mode resonators (WGMRs) as shown in earlier works[46-49]. However, all the three coupled light waves must be on the resonance simultaneously in a WGMR OPO, and makes the phase matching complicate for comb generation.

Here we revisit the concept of the box resonator, and realize its high-Q fabrication at MIR wavelength, for the first $\chi^{(2)}$ mid-infrared optical frequency comb generation with continuous wave near-infrared pump. This optical superlattice box resonator (OSBR) is fabricated by fine polishing over a periodically poled lithium niobate crystal, down to the size of 0.1 *mm*×0.5 *mm* ×44.1 *mm*. Near-material-limited quality factor of $2.0\times10^7$ is measured at 2060 *nm*, with small anomalous dispersion of 19.5 *ps/(nm·km)*. By pumping at 1030 *nm*, the comb span exceeds over 250 *nm*, with maximum output power of 370 *mW* and 9.0 % peak conversion efficiency. 1.566 *GHz* comb beat note is measured with a narrow linewidth of 6.1 *kHz*, and a low comb noise is revealed with consistently revival temporal waveforms.

A perfect electromagnetic resonator can be formed by a rectangular boundary over a dielectric medium, and such box resonator is a popular resonator model in the text books [50], as an example shown in Fig. 1a. For optical wave, we use this old concept to realize a new type of high-Q resonator by fine mechanical polishing. In this configuration (Fig. 1b), all the six box surfaces are polished for a transverse waveguide confinement by total internal reflection, and the longitudinal confinement can be achieved by wavelength selective optical coatings. Here the coating is chosen for doubly resonant at MIR signal and idler wavelengths and single-pass for pump, which greatly simplifies the phase matching in the resonator. Fig. 1c shows the fabrication process, and it starts from the optical superlattice fabrication on a 3-inch MgO-doped lithium niobate wafer with our whole wafer poling technique. Both wafer surfaces are then polished down to the thickness of 100 *μm* with sub-nanometer roughness and coated with 110 *nm* $Ta_2O_5$ cladding layers to minimize the MIR absorption, which are followed by 200 *nm* $SiO_2$ layers for crystal bounding. Lithium tantalite (LT) substrates are used to sandwich the optical superlattice wafer for easy handling and better thermal management. After slicing and fine polishing, the OSBR can be fabricated with coating on the two small end faces with 99 % and 99.8 % reflection, respectively, at around 2060 *nm* and anti-reflection at 1030 *nm*. Each resonator is mounted in a thermal conductive metal housing for precise temperature control.

The poling period of the OSBR is 31.2 *μm* for the QPM optical parametric down conversion process from 1030 *nm* to 2060 *nm*. Such design enables the parametric light generation close to the material zero-dispersion point and with small anomalous dispersion, which is good for the comb generation. With a relatively large cross-section size of 0.1 *mm*×0.5 *mm*, the OSBR has big number of transverse modes, but simulation shows that the low-order modes have similar anomalous dispersion that is close to the bulk material (Fig. 2a, for details, see supplementary information I). To characterize the Q factor and dispersion of OSBR, we build a tunable MIR source by difference frequency generation (DFG) between a CW Ti-sapphire laser (M Squared SolsTiS) and an erbium-doped fiber amplified tunable laser (Santec TSL-710) (For details, see supplementary information I). The scan result is shown in Fig. 2b, only four sets of transverse modes, TM00, TM01, TM02 and TM03 can be identified. The fundamental mode has the highest Q and thus transmission, with free spectral range (FSR) of about 1.566 *GHz*, which agrees well with the 44.1 *mm* cavity length. The inset shows the zoom-in of the resonance, and the linewidth is fitted to be 7.2 *MHz* full width half maximum (FWHM), which corresponds to a Q factor of over $2.0 \times 10^7$. We focus on the fundamental mode for the following study.

For the MIR OFC generation, the pump source is a continuously ytterbium-doped fiber amplifier (YDFA) seeded by a tunable laser (Toptica CTL 1050), as shown in Fig. 2c. A small portion of pump is reflected by 4 % beam splitter for frequency and power stabilization. A wavelength meter (HighFinesse WS-U) is used to lock the seed laser and achieves a long-term stability within 1.5 *MHz* at around 1030 *nm* wavelength. The intensity noise from the YDFA is suppressed using a PID feedback loop, to achieve reasonable thermal stability on the resonances of the OSBR. The pump beam is then directed through a set of cylindrical lens to match the mode profile of the resonator. The OSBR is placed on a double-enclosed metal mount, which is temperature stabilized with Peltier cooler. The internal metal enclosure is also temperature stabilized and isolated from the external environment by an external enclosure. By using high performance temperature controllers, sub-milliKelvin temperature stability can be achieved for the OSBR for stable doubly-resonance with the pump. A pump rejection filter is used to remove the residue pump light after the resonator. Despite of the large mode size, the optical parametric oscillation is observed with a low pump power threshold of about 0.35 *W*. We measure the power of the parametric light as a function of pump, and the result is shown in Fig. 2d. The maximum output power of 0.37 *W* can be measured at 9.1 *W* pump, and the peak conversion efficiency of 9.0 % can be calculated at 1.0 *W* pump.

We couple the parametric light into a single mode fiber to further study on its spectral and temporal features. We first set the OSBR temperature for degenerate QPM with 1030 nm pump. By increasing the pump power, comb-like spectrum can be captured with a grating imaging spectrometer (Princeton Instruments SP-2500). At pump power of 9.1 *W*, we observe the MIR OFC with a span of about 20 *nm* around 2060 *nm* (Fig. 3a). By slightly reducing the temperature, the QPM condition can be shifted to the nondegenerate wavelengths. With proper detuning the pump frequency relative the OSBR resonances, the comb span can be further extended. In experiment, we observe a maximum spectral span of about 250 *nm* (Fig. 3b). Both spectrums agree well with our simulation shown in Figs. 3c and 3d.

The temporal behavior of the MIR OFC is further studied using a fast InGaAs detector. With proper pump detuning, revival temporal waveforms can be observed on a fast oscilloscope (Textronics MSO 73304DX) at both degenerate and nondegenerate points, as shown in Figs. 4a and 4b. For a better characterization of the noise feature, we plot them stacked together in the right figures, within one repetition time window of about 0.639 *ns*, they match well with each other, with a standard deviation of less than 3 %. An electronic spectral analyzer (R&S FSVA30) is used to measure the RF spectrum in the nondegenerate case, with results shown in Fig. 4c. Clean beat note peaks can be captured at integer number times of the repetition frequency at 1.566 *GHz*, which is close to the cavity FSR of the 44.1 *mm* length of the box resonator. Fig. 4d shows the zoom-in around the repetition frequency, and this RF peak is fitted to be with a narrow FWHM linewidth of 6.1 *kHz*. The above results prove a stable and low-noise MIR OFC generation from our OSBR.

In summary, we have demonstrated the first mid-infrared optical frequency comb generation from $\chi^{(2)}$ optical superlattice box resonator. This new OSBR platform allows the access to near-material-limited quality factor of $2.0 \times 10^7$ for lithium niobate with coating-configurable resonance design. Here we use it for doubly-resonance at signal and idler light around 2060 *nm*, where a small anomalous dispersion can be achieved with artificial quasi-phase matching using domain engineering. As a result, MIR OFC can be generated with 250 *nm* span and 370 *mW* output power. Repetition rate of 1.566 *GHz* is measured by fast detectors with beat note linewidth of 6.1 *kHz*, which may be further improved with better stability control of the OSBR and pump laser. Consistent temporal recurrent is measured and shows a low comb noise. In this study, we focus on the simplest optical superlattice with periodical poling structures. More complex structures like chirped structure can be applied to the OSBR for octave spanning OFC generation, and nonlinear dispersion control is possible with proper chirping to manipulate comb dynamics towards soliton comb generation. Therefore, such $\chi^{(2)}$ resonators enable extra flexibilities for MIR OFC generation, which are important for high resolution spectroscopy, sensing and ranging applications.

**Acknowledgements**

The authors acknowledge Prof. Shuwei Huang, Prof. Xiaoshun Jiang, Ming Nie, and Bowen Li, for fruitful discussions and suggestions regarding the manuscript. This work is supported Ministry of Science and Technology of the People's Republic of China (No. 2017YFA0303700), the National Young 1,000 Talent Plan; National Natural Science Foundation of China (No. 51893086014, No. 11674169, No. 11621091, No. 91321312).

**Author contributions**

Kunpeng Jia, Xiaohan Wang and Zhenda Xie conceived the original idea and designed the experiment. Kunpeng Jia, Xiaohan Wang, Liyun Hao and Jian Guo fabricated the sample and performed the measurement. Xin Ni, Huaying Liu and Xinjie Lv provided the theory and simulation. Zhenda Xie and Shining Zhu supervised the whole work, and all authors contributed on the manuscript preparation.

**Competing interests**

The authors declare no competing interests.

**Figures**

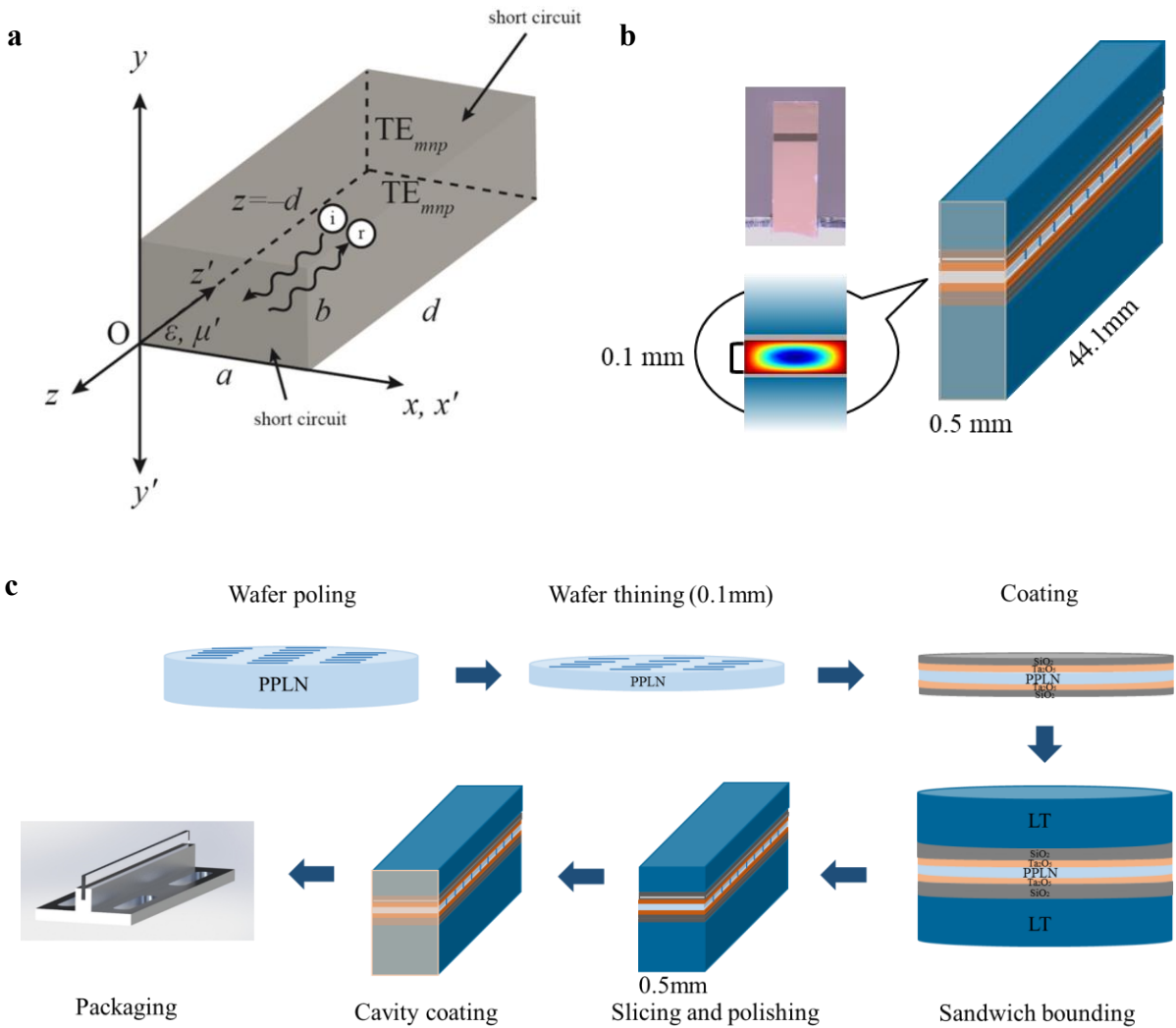

**Figure 1 | The schematic of an optical superlattice box resonator. a,** Textbook image of electromagnetic wave confinement with rectangular boundary. **b,** Structure of the OSBR. The optical superlattice "box" is sandwiched by lithium tantalite (LT) substrates, with the effective size of $0.5 \times 0.1 \times 44.1$ $mm^3$. **c,** Fabrication process of the OSBR. The optical superlattice wafer is thinned to 0.1 $mm$ by mechanical polishing and then sandwiched by lithium tantalite (LT) substrates with tantalum pentoxide and silicon dioxide film as buffer layers. Then it is sliced and precisely polished for all the surfaces. Optical coatings are applied with wavelength selectivity for doubly resonances at MIR wavelength. S1: AR @ 1030 $nm$, R=99.8 % @ 2060 $nm$; S2: AR @ 1030 $nm$, R=99 % @ 2060 $nm$. Finally, the OSBR is mounted in thermal conductive metal housing for thermal control.

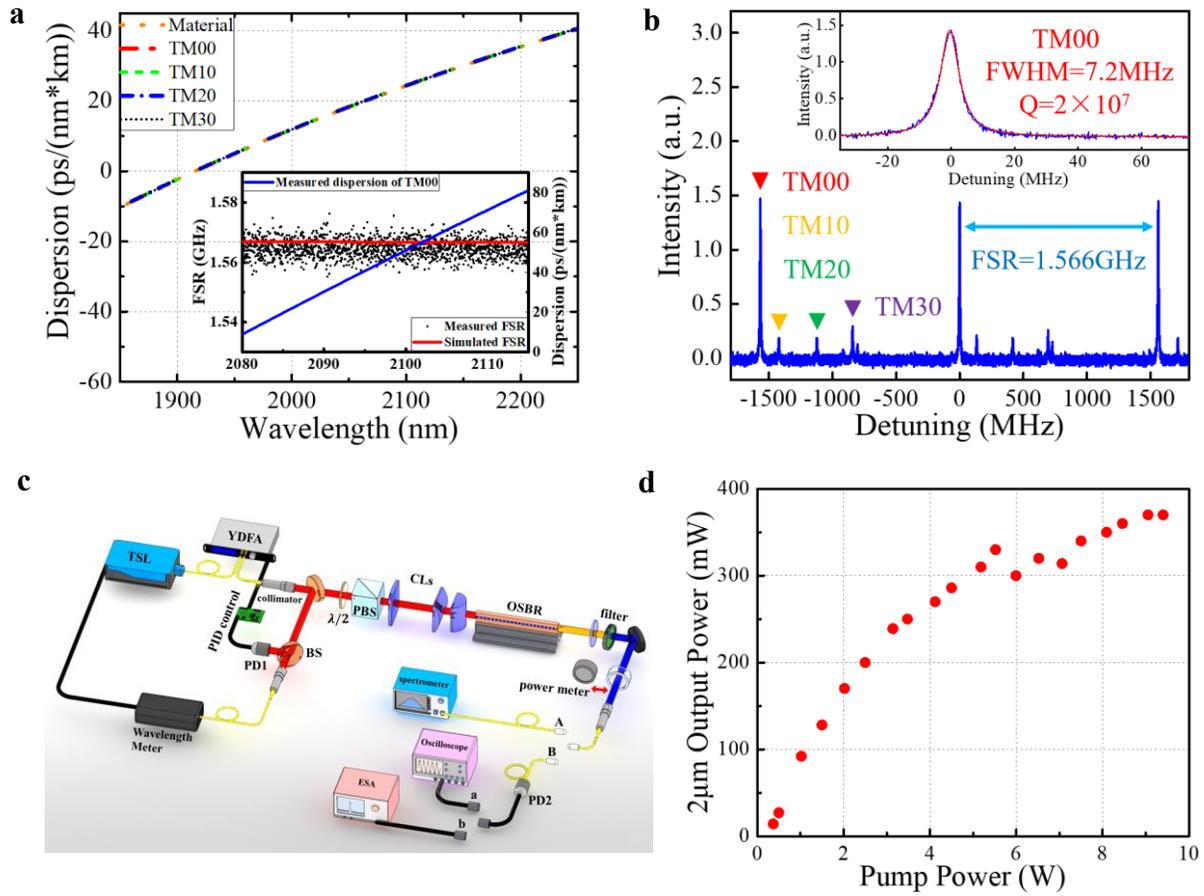

**Figure 2 | Dispersion characteristic of the OSBR and the experimental setup. a,** Small anomalous dispersion is simulated for OSBR around 2060 *nm*. Due to the relatively large mode field area of OSBR, the dispersion of waveguide modes is very close to the bulk material dispersion. Inset: Measured FSRs and simulated dispersion of TM00 mode as a function of wavelength. A near-to-zero dispersion can be fitted, and it has small deviation from theoretical values because of the limited system resolution. **b,** OSBR transmission by frequency scan using our DFG MIR source. Resonances of four mode families (TM00, 10, 20, 30) can be observed. The insert is a zoom-in picture of one transmission peak of TM00 mode with Q factor of $2.0 \times 10^7$. **c,** Experiment setup for MIR OFC generation. TSL: tunable semiconductor laser, YDFA: Ytterbium-Doped Fiber Amplifier, CLs: cylindrical lens, PD: photo detector, BS: beam splitter, PBS: polarization beam splitter. ESA: electronic spectrum analyzer. **d,** The output power of the MIR OFC as a function of the pump power. The maximum power exceeds 0.37 *W* with an OPO threshold of 0.35 *W* and a maximum conversion efficiency of 9.0 %.

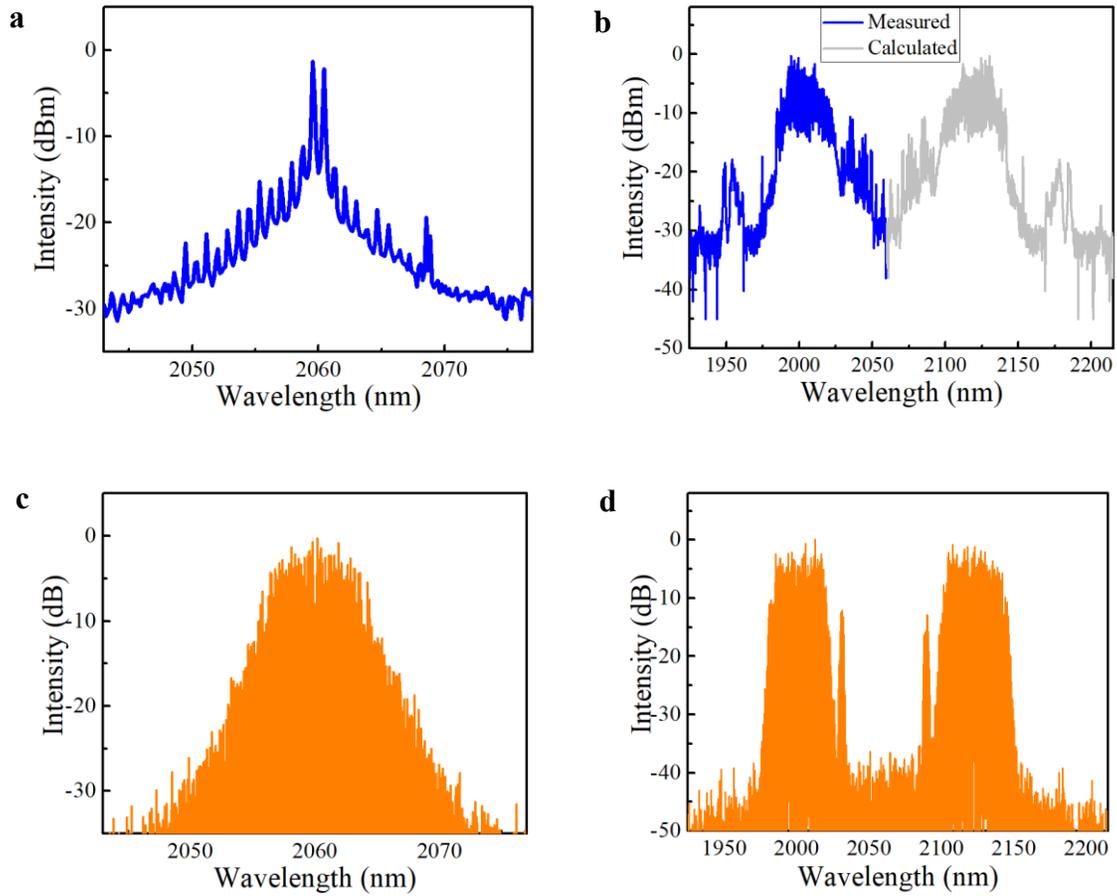

**Figure 3 | Mid-infrared frequency comb generation. a,** MIR OFC spectrum at degenerate QPM point with 1030 *nm* pump at power of 9.1 *W*. Comb span of about 20 *nm* can be measured. **b,** MIR OFC spectrum at nondegenerate QPM point. Only the short wavelength spectrum (blue) is measured because of the span limit of our spectrometer, and the corresponding long wavelength spectrum (grey) is calculated according to energy conservation. The comb span exceeds over 250 *nm*. **c, d,** Simulated spectrum at degenerate point (Fig. c) and nondegenerate point (Fig. d).

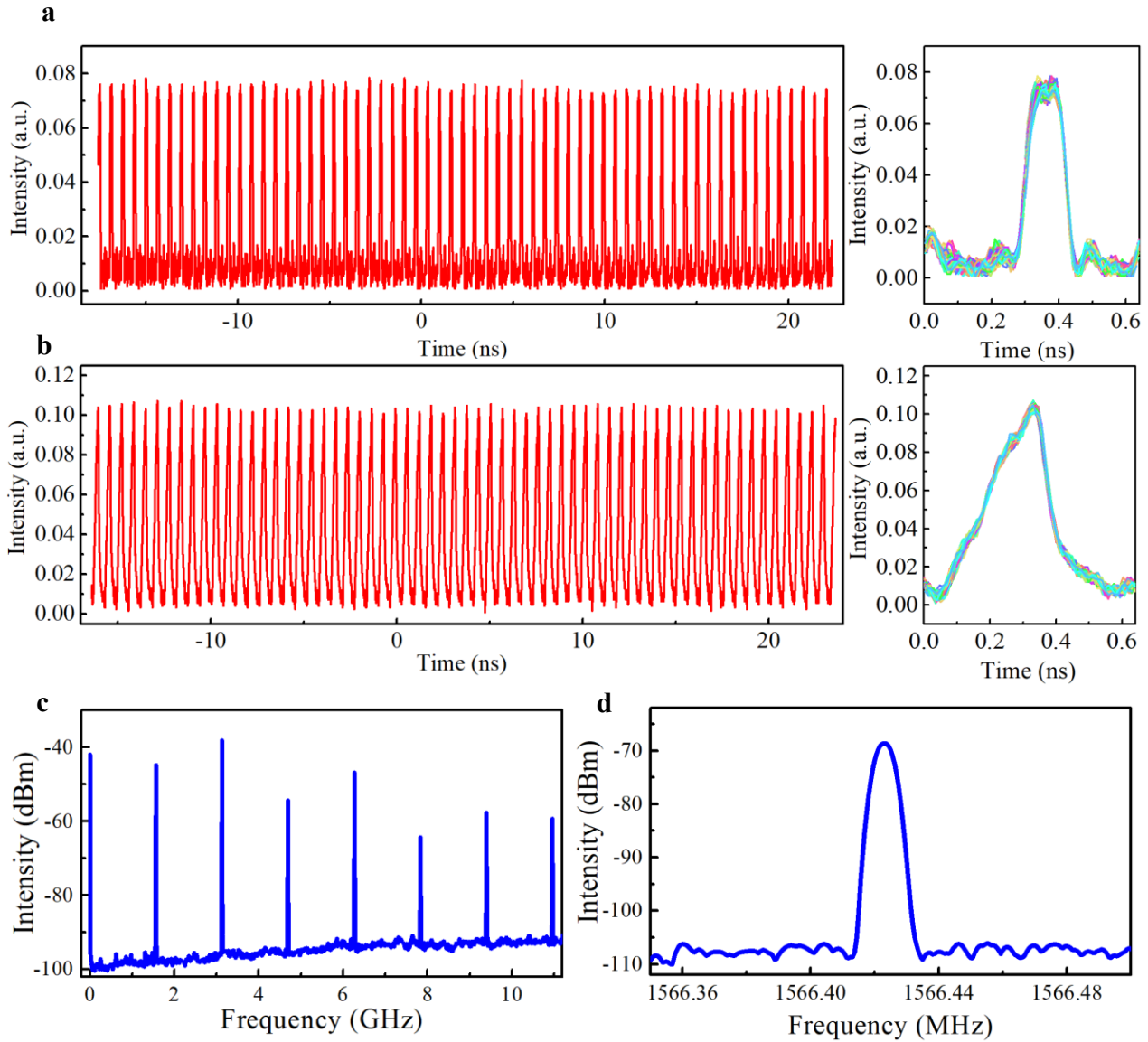

**Figure 4 | Temporal waveform and electronic spectrum measurement. a, b** The time domain waveform of 62 repetitions at degenerate and nondegenerate QPM conditions. These waveforms are stack-plotted within one repetition in the right figures, which show good reproducibility. **c,** The RF beating spectrum of the photo current. **d,** Zoom-in of RF spectrum around repetition frequency at 1.5664 *GHz*. A linewidth of 6.1 *kHz* can be fitted (resolution bandwidth, RBW = 5 *kHz*).

# Supplementary Information for

# Mid-infrared optical frequency comb generation from a $\chi^{(2)}$ optical superlattice box resonator

Kunpeng Jia[+], Xiaohan Wang[+], Xin Ni, Huaying Liu, Liyun Hao, Jian Guo, Jian Ning, Gang Zhao, Xinjie Lv, Zhenda Xie*, and Shining Zhu*

*National Laboratory of Solid State Microstructures, School of Electronic Science and Engineering, College of Engineering and Applied Sciences, and School of Physics, Nanjing University, Nanjing 210093, China*

[+]*These authors contributed equally to this work.*

* E-mail: xiezhenda@nju.edu.cn,
zhusn@nju.edu.cn

**I. Mode dispersion in the OSBR**

We use the finite element analysis method to simulate the mode profile and dispersion for the OSBR. The effective cross-section size is 0.1 $mm \times$ 0.5 $mm$. Fig. S1a shows the profile of the TM00, TM10, TM20 and TM30 modes. Because of the relatively large cross-section, these low-order modes have their effective indexes very close to each other and that of the bulk PPLN, as listed in Table S1. For a Fabray-Perot type resonator, the resonance condition can be calculated as follows

$$n_{effi} L = N \frac{\lambda_i}{2} = N \frac{c}{2 f_i} \tag{1}$$

where $N$ is the number of the wave node inside the resonator, $\lambda_i$, $f_i$, $n_{effi}$, $L$ are the wavelength in vacuum, frequency, the effective mode index of the TM$i$0 mode, and cavity length, respectively. Therefore, the relative detuning of $i$th mode family to TM00 mode can be calculated by Eq. (2):

$$\Delta f_i = f_i - f_0 = N \frac{c}{2 n_{effi} L} - N \frac{c}{2 n_{eff0} L} = f_0 \frac{n_{effi} - n_{eff0}}{n_{effi}} \tag{2}$$

Here the mode indexes are so close to each other that $\Delta f_i$ are smaller compared to the FSR. Therefore, there is no mode crossing for the low-order modes. We simulate the mode indexes of these transverse modes and calculated their relative detuning to fundamental mode around 2100 $nm$, which matches well with experimental results as shown in Fig. S1b and Table S1.

Fig. S1c shows the schematic of the dispersion measurement setup. To get a laser with wide and mode-hop-free tuning range at MIR wavelength, we build a difference frequency generation (DFG) setup between a CW Ti-sapphire laser (M Squared SolsTiS) and an erbium-doped fiber amplified (EDFA) tunable semiconductor laser (TSL, Santec TSL-710), using a bulk PPLN crystal. Pumping at 890.47 $nm$, we sweep the wavelength of TSL through 20 $nm$ range with a tuning rate of 100 $nm/s$, which corresponds to 37 $nm$ range at MIR wavelength. 1.5 % of the TSL output is directed into a hydrogen cyanide (HCN) gas cell for wavelength calibration according to the absorption features, and another 1.5 % of the TSL light is directed through a fiber Mach-Zehnder interferometer (MZI) to generate a clock signal from PD3 to correct the nonlinearity of the scan. The MZI is built with 15.8 *m* unbalanced path lengths, which translates to a 12.6 *MHz* optical frequency sampling resolution. 5 *m* long dispersion compensating fiber (DCF38) is used in MZI for compensating the dispersion in the long arm. The DFG MIR beam is then coupled into OSBR and the transmission intensity is detected by photodetector (PD1) as shown in Figs. S2a and S2b. Therefore, wavelength dependent FSR and dispersion of OSBR can be calculated from the above measurement, with result shown in Fig. 2a in the main text.

**II. Simulation of the comb generation dynamics in the OSBR**

In this simulation, we use the split-step Fourier method for the MIR OFC generation[1], which is a pseudo-spectral numerical method to solve the nonlinear Schrödinger equation. In the split-step Fourier method, the OPO process is split into small alternating steps, with nonlinear interaction and linear propagation, respectively. While the nonlinear step is calculated in time domain, the linear step is calculated in frequency domain for maximum simulation efficiency, and Fourier transform is performed between steps for continuous simulation. Here in this 44.1 *mm* long OSBR, we can have a maximum bandwidth on the order of 10 *THz*, and thus sub 100 *fs* temporal resolution is used in the simulation.

Similar to the experiment described in the main text, both degenerate and nondegenerate QPM cases are simulated in this doubly resonant OSBR. We scan the pump detuning within ±2.5 *MHz* relative to the resonances and search for lowest comb noise, which is characterized by the cross correlation between different repetition. The results are shown in Figs. S3a and S3b. The cross correlation has small deviations under $5.8\times10^{-6}$ and 0.67 % to unitary at 0.2 and 0.46 *MHz* detuning, in the degenerate and nondegenerate cases, respectively. At both cases with above detunings, we simulate the spectrum evolution for the MIR OFC build up in the time domain. The results are shown in Figs. S3c and S3d, which shows low spectral noises.

The physics of the coherent MIR OFC generation from OSBR can be explained as result of cascad $\chi^{(2)}$ process as shown in Fig. S4. Assuming adjacent comb line pairs $\omega \pm n\Omega$ and $\omega \pm (n+1)\Omega$ ($\Omega$ stands for the comb line spacing) are generated via an OPO process pumped by a beam at frequency $2\omega$, the cavity enhanced sum frequency generation (SFG) process between $\omega - n\Omega$ and $\omega + (n+1)\Omega$ is within the QPM bandwidth to generate $2\omega + \Omega$, and DFG happens between this SFG light and comb line $\omega - (n+1)\Omega$ with output frequency of $\omega + (n+2)\Omega$. This new comb line can seed into the following optical parametric amplification (OPA) process: $2\omega \rightarrow [\omega + (n+2)\Omega] + [\omega - (n+2)\Omega]$, generating new sideband around frequency $\omega$ and cascading over the whole bandwidth.

**Supplementary Reference**

**Supplementary Table**

Table S1 | Mode indexes and frequency detunings at 2.1 *μm* wavelength (23 ℃)

|          | Simulated mode index | Simulated frequency detuning to TM00 (*GHz*) | Measured frequency detuning to TM00 (*GHz*) |
|----------|---------------------|---------------------------------------------|---------------------------------------------|
| Material | 2.11562872281008    |                                             |                                             |
| TM00     | 2.11560391895594    | 0                                           | 0                                           |
| TM10     | 2.11560158799612    | 0.157                                       | 0.142                                       |
| TM20     | 2.11559770303373    | 0.420                                       | 0.436                                       |
| TM30     | 2.11559226403033    | 0.787                                       | 0.720                                       |

**Supplementary Figures**

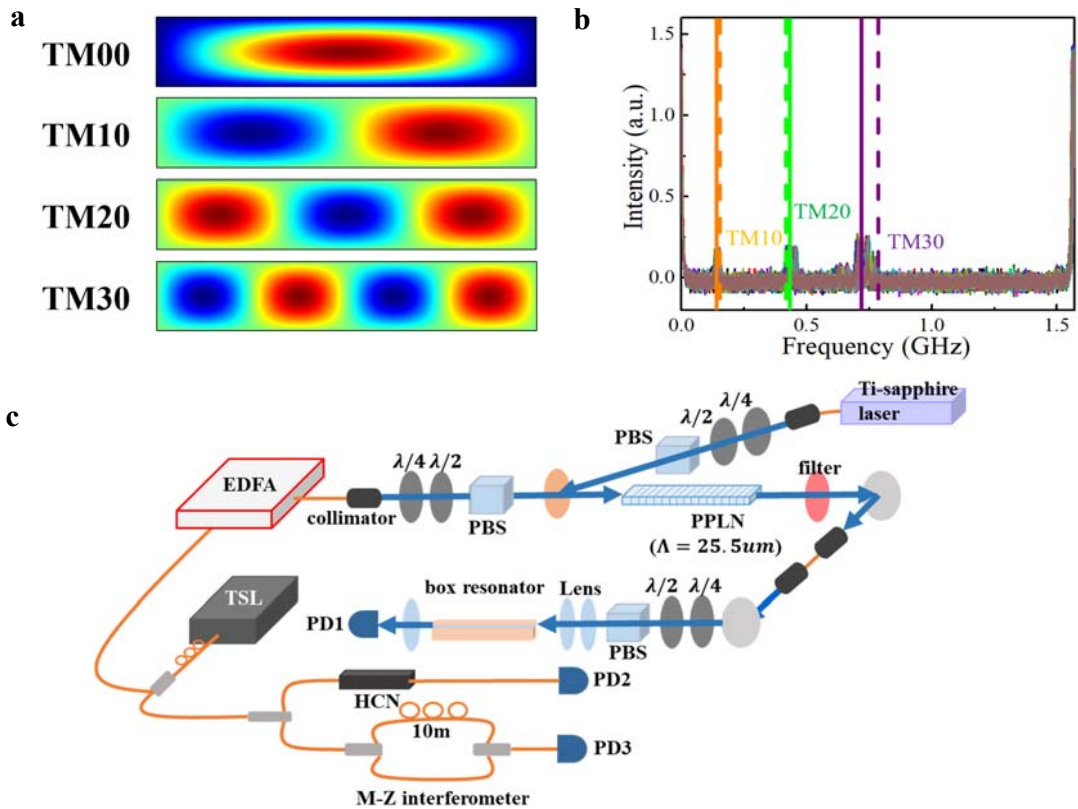

**Figure S1 | The mode profile and dispersion of OSBR. a,** Simulated mode profiles of TM00, TM10, TM20 and TM30. **b,** Stacked transmission plots of 20 FSRs based on fundamental mode around 2.1 $\mu m$. The solid lines mark the experimental position of transverse modes TM10, TM20 and TM30 in one FSR range, respectively, which shows a good agreement with the simulation results plotted by dashed lines. **c,** Dispersion measurement setup. TSL, tunable semiconductor laser; EDFA, erbium-doped optical fiber amplifier; PBS, polarization beam splitter; PD, photodetector.

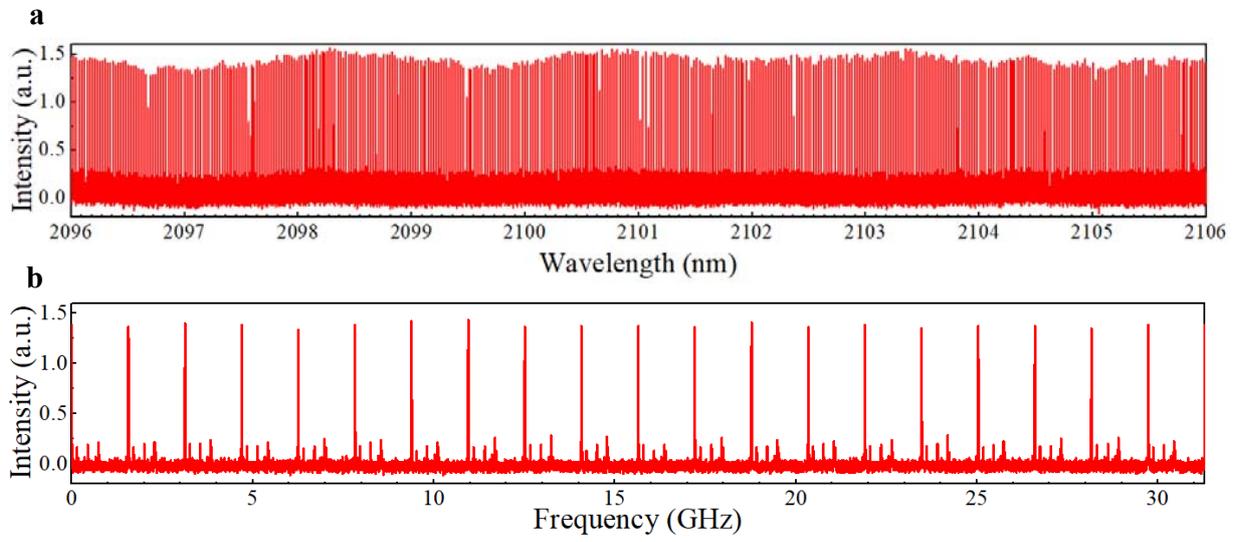

**Figure S2 | MIR transmission intensity of OSBR a,** Transmission of the cavity modes in 10 *nm* range. **b,** Zoom-in of the transmission of 20 FSRs around 2.1 *μm*.

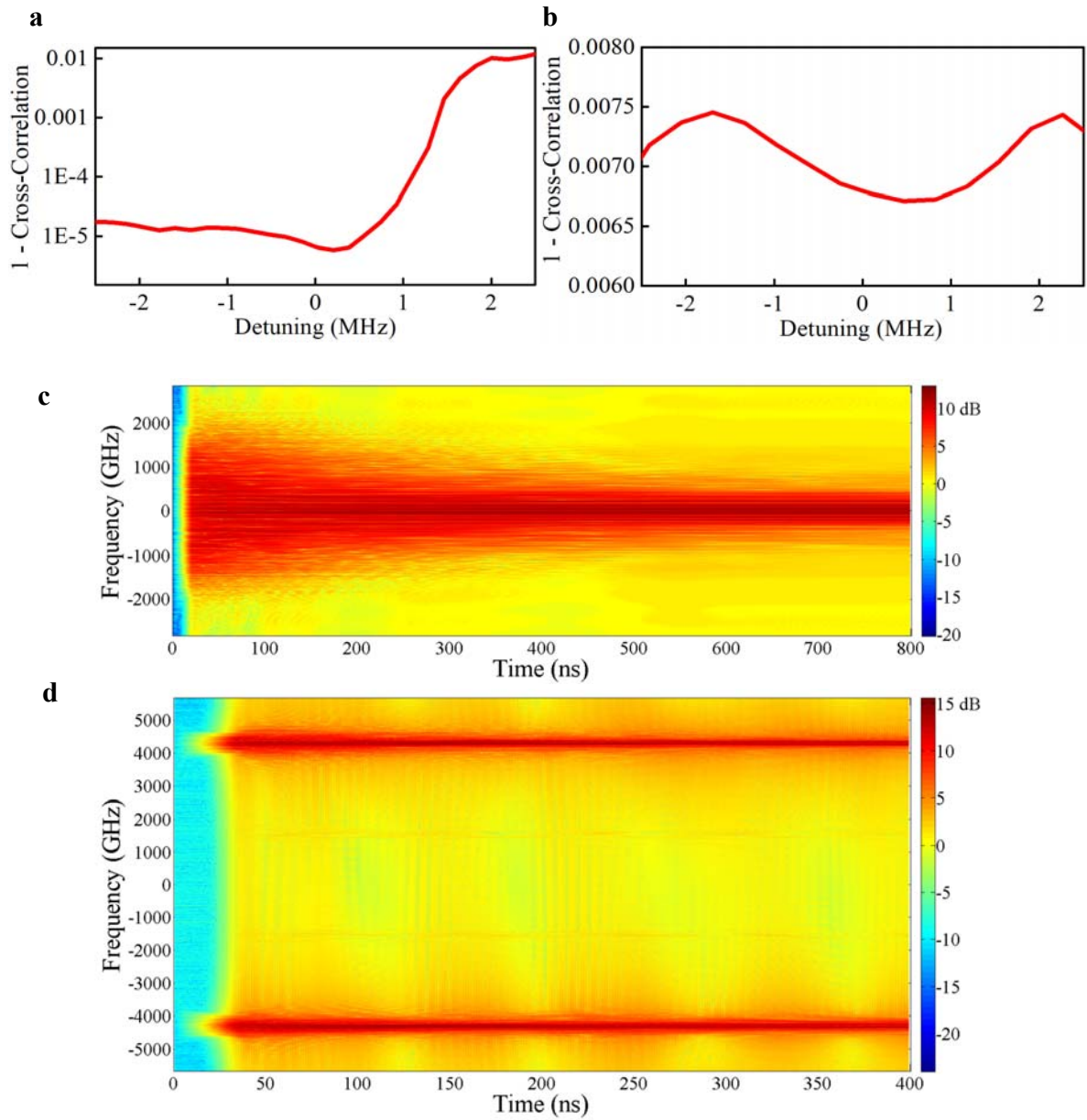

**Figure S3 | Simulation of the comb generation at degenerate and nondegenerate points. a, b,** The noise level as a function of pump detuning at degenerate and nondegenerate point, respectively. **c, d** Spectrum evolution for the MIR OFC build up in the time domain at degenerate and nondegenerate point, respectively.

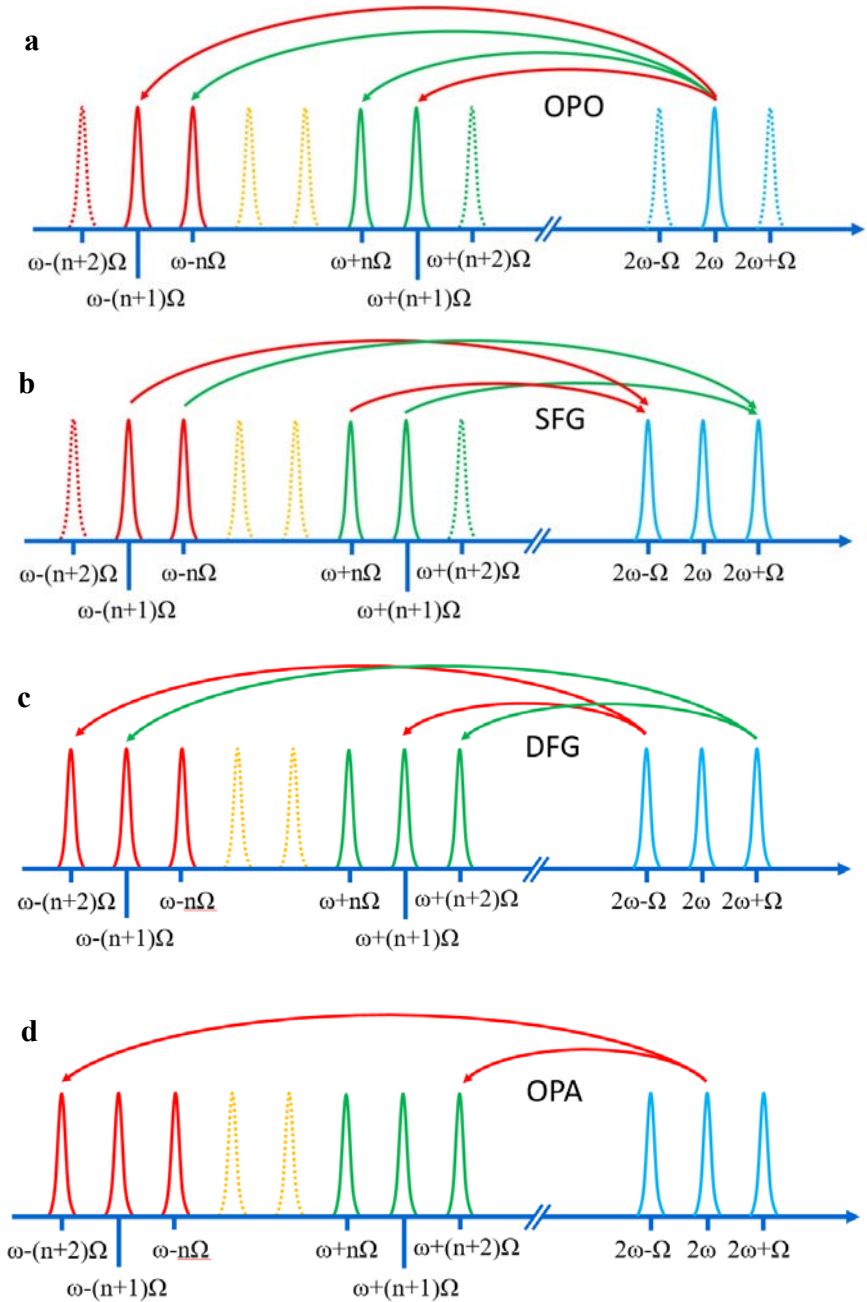

**Figure S4 | MIR OFC generation process. a,** Optical parametric oscillation (OPO). **b,** Sum frequency generation (SFG). **c,** difference frequency generation (DFG). **d,** Optical parametric amplification (OPA).